\documentclass[12pt]{article}
\usepackage{amsmath}
\usepackage{graphicx}%
\usepackage{amsfonts}%
\usepackage{amssymb}
\usepackage{epsfig}

\newcommand{\bl}{\pmb{l}}
\newcommand{\bk}{\pmb{k}}

\newcommand{\be}{\begin{equation}}
\newcommand{\ee}{\end{equation}}

\newcommand{\bea}{\begin{eqnarray}}
\newcommand{\eea}{\end{eqnarray}}
\newcommand{\cM}{{\cal M}}

\usepackage{graphicx}

\begin{document}

\title{\Large \bf 
In search of the QCD odderon in exclusive $J/\psi$ and $\Upsilon$ hadroproduction
}
\author{\large Lech Szymanowski$^1$ \bigskip \\
{\it  $^1$~CPhT, Ecole Polytechnique, CNRS, 91128 Palaiseau, France} \\ {\it
and Soltan Institute for Nuclear Studies, Warsaw, Poland}}

\maketitle


\begin{center}
{\bf Abstract}\\
\medskip
Phenomenological studies of odderon effects are shortly reviewed. Special emphasis is devoted to  a  recent
 study of the exclusive production of $J/\psi$ or $\Upsilon$ in $pp$ and $\bar pp$ collisions, where the meson emerges 
from the pomeron--odderon and the pomeron--photon fusion. 
\end{center}

\section{Introduction} \label{s1}

Hadronic elastic processes occuring at high energies and at small momentum transfers are described by exchanges of either the pomeron with even charge parity or by its odd charge parity conterpart, the odderon. In the framework of the  perturbative QCD, the pomeron and the odderon are described in the Born approximation by exchanges of, respectively,   two and three gluons. The higher order effects of interation between exchanged gluons are summarized in the leading logarithmic approximation (LLA) by the BFKL equation \cite{BFKL} for the pomeron and by the BKP equation \cite{BKP}
for the odderon.

The concept of the odderon  was introduced long ago \cite{Lukaszuk}; from the point of view of general principles based on the analyticity and the unitarity of the $S$-matrix one could 
naively expect that its exchange should 
lead  to effects of comparable magnitudes to those of the much well known pomeron. Nevertheless, till now it is hard to observe fully convincing effects of the odderon which somehow escapes experimental verification and remains some mystery. On the other hand, the attempts to understand in details the QCD dynamics of interacting gluons forming pomeron and odderon has lead to numerous theoretical developments showing its connections with  exactly integrable
models of the non-compact spin chains \cite{integrability}, the gauge theory/string duality \cite{duality} or to the dipole model \cite{dipolemodel}.

\section{Phenomenology with the odderon} \label{s1}
\subsection{Elastic $pp$ and $p\bar p$ scattering}

The most convincing trace of the odderon effects would be a discovery of the difference between the differential cross sections for proton-proton and antiproton-proton at high energies and at small momentum transfers. Such an analysis of the difference between $(d\sigma/dt)_{\bar p p}$ and $(d\sigma/dt)_{ p p}$ in the dip shoulder region $1.1 < |t| <1.5\,$GeV$^2$ at $\sqrt{s}=52.8\,$GeV was already performed  at the ISR at 1985 and led to some evidence for the presence of the odderon \cite{ISR}. Unfortunately, due to the closure of the ISR, the obtained results are not fully convincing. The recent study \cite{Avila} discusses how to find the odderon at RHIC and at LHC. At RHIC
the experiment can be done with the STAR detector in combination with the existing UA4/2 data.  Fig.~1a shows the
predictions for  $(d\sigma/dt)_{\bar p p}$ and $(d\sigma/dt)_{ p p}$;  the spectacular oscillations in the difference of cross sections $\Delta\left( \frac{d\sigma}{dt} \right)(s,t)\equiv 
| \left( \frac{d\sigma}{dt} \right)^{p\bar p}(s,t) - 
\left( \frac{d\sigma}{dt} \right)^{pp}(s,t) |$ are shown in Fig.~1b. The corresponding predictions for the LHC lead in particular to the values $\sigma_T(\sqrt{s}=14\,\mbox{TeV})=123.32\,$mb and $\Delta\sigma(\sqrt{s}=14\,\mbox{TeV})=(\sigma^{\bar p p}_T-\sigma^{ p p}_T)(\sqrt{s}=14\,\mbox{TeV})=-3.92\,$mb.

\begin{figure}[h]
\center{
\includegraphics[width=2in]{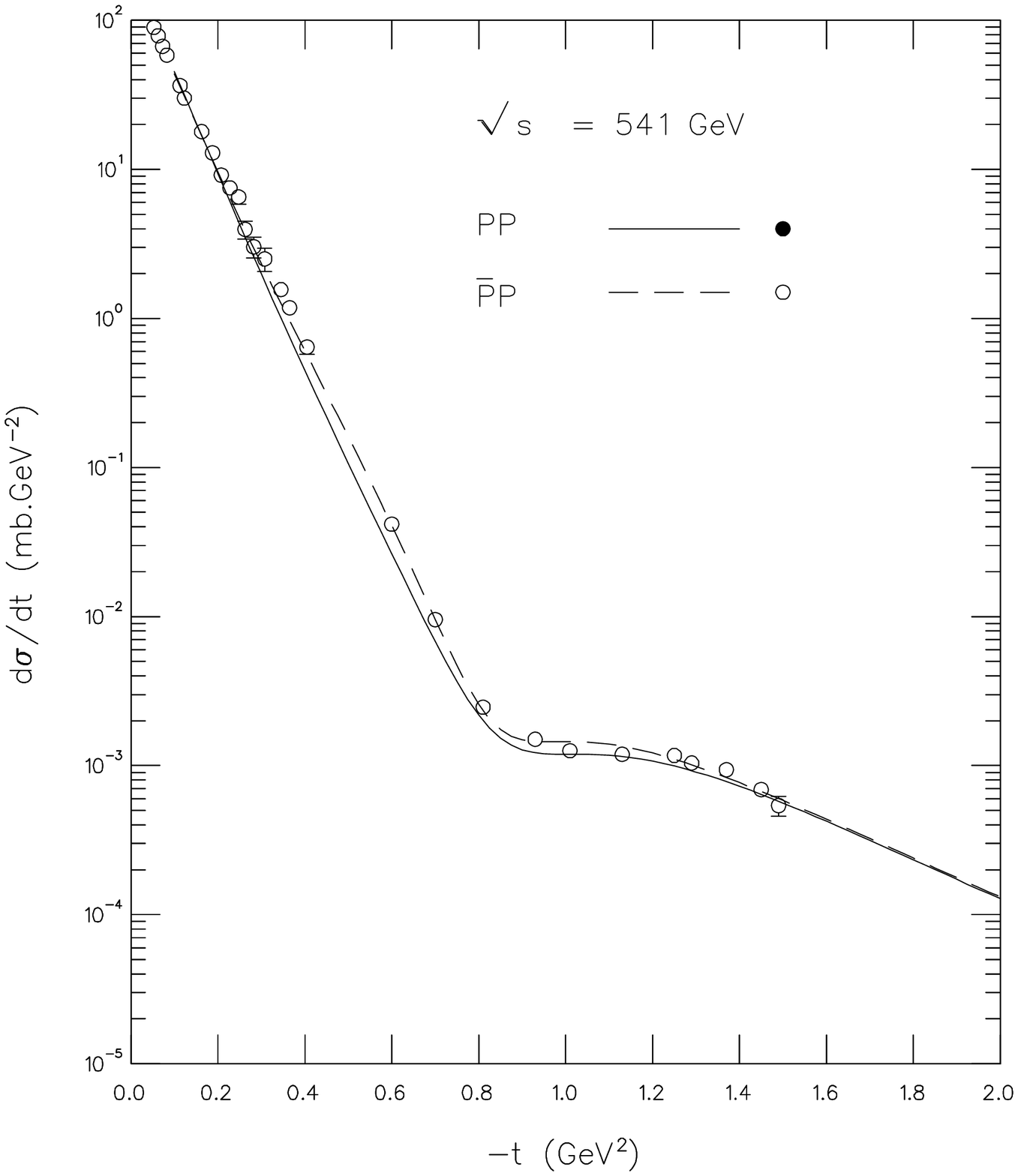}
\hspace{2cm}
\includegraphics[width=2in]{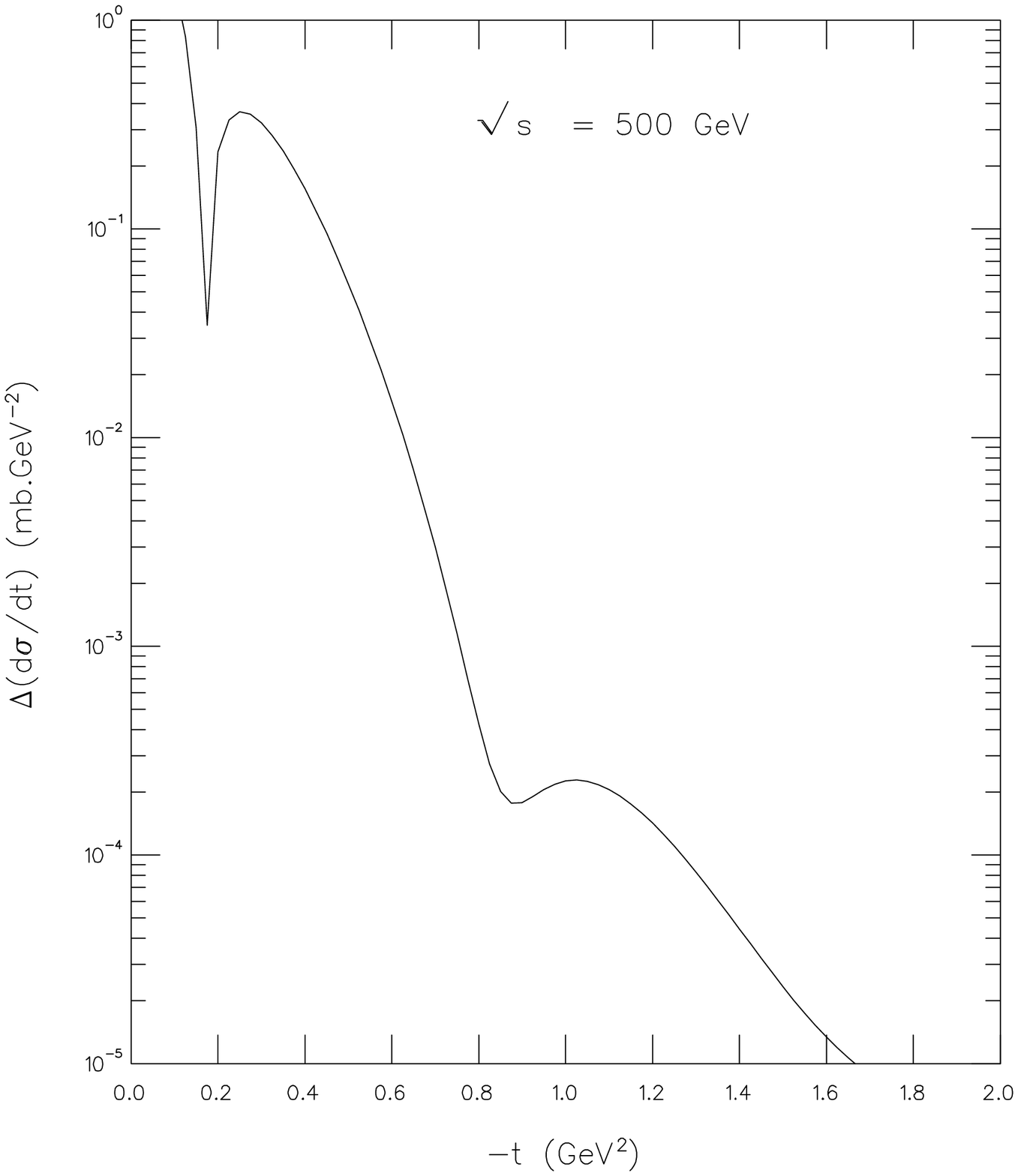}
\hspace{2.5cm}
(a)$\hspace{8cm}$(b)
\caption{ (a) $pp$ and $\bar p p$ predictions of \cite{Avila} with the UA4/2 $\bar p p$ points, (b) Predictions for \hspace*{2cm}$\Delta(d\sigma/dt)$.   }
}
\end{figure}

\subsection{Exclusive meson electroproduction}

An interesting possibility to search for the odderon effects is the diffractive production of pseudoscalar or tensor mesons $\gamma^{(*)} p \to M\, p$ shown in Fig.~2a. In particular, if meson M is an $\eta_c$ pseudoscalar charmonium,  the mass of the $c-$quark supplies a hard scale which justifies the perturbative description of the odderon
\cite{KwiecinskiMotyka,etac}. 
It is natural to expect that in hard processes the effects of odderon exchanges - being suppressed by  an additional power of the strong coupling constant $\alpha_s$ - are smaller than similar contributions due to pomeron exchange, e.g. when $M=J/\psi$. Unfortunately, the obtained predictions, as those presented in Fig.~2b \cite{etac} are very small 
with respect to experimental possibilities.
 The situation is not  significantly improved by taking into account the effects of the BKP evolution, which leads to an  enhancement factor of the order 5 \cite{etacBKP}.

\begin{figure}[h]
\epsfxsize=3.5cm
\begin{center}
\includegraphics[width=2.3in]{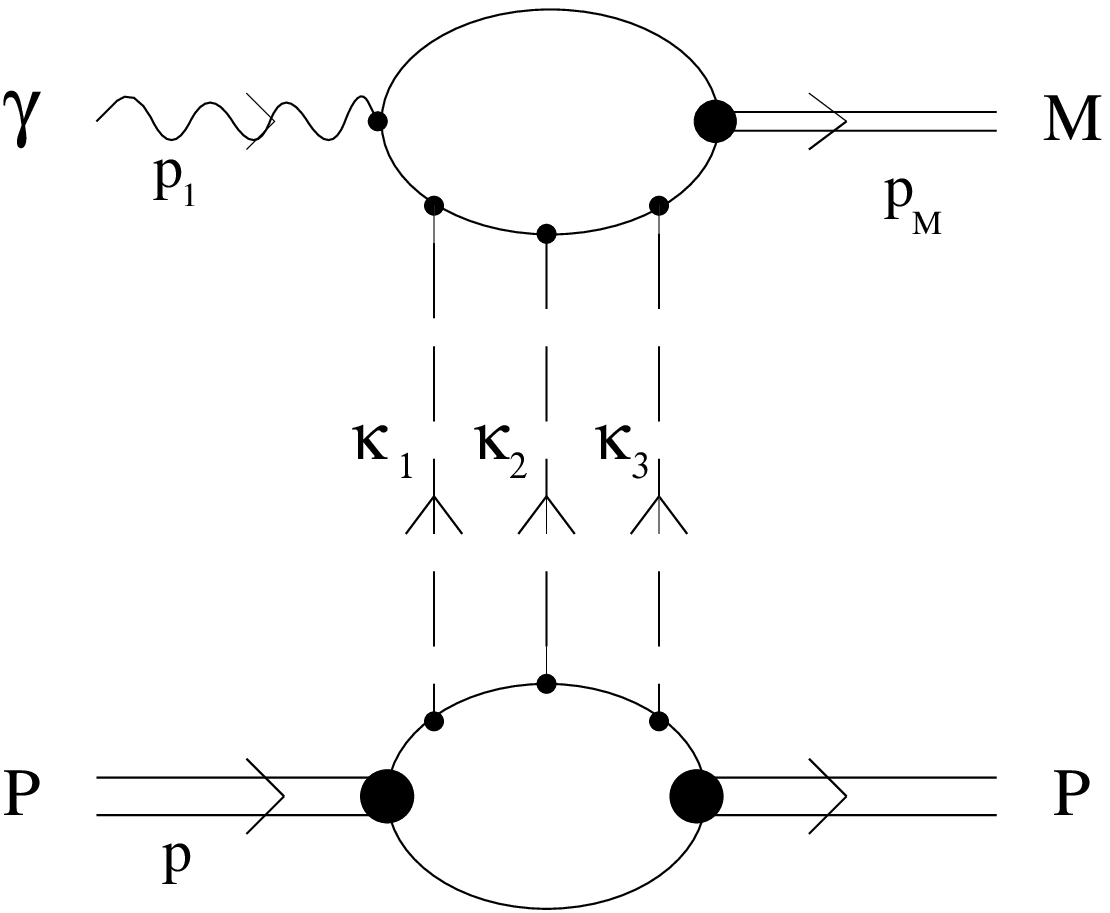}
\hspace{1.5cm}
\includegraphics[width=2.0in]{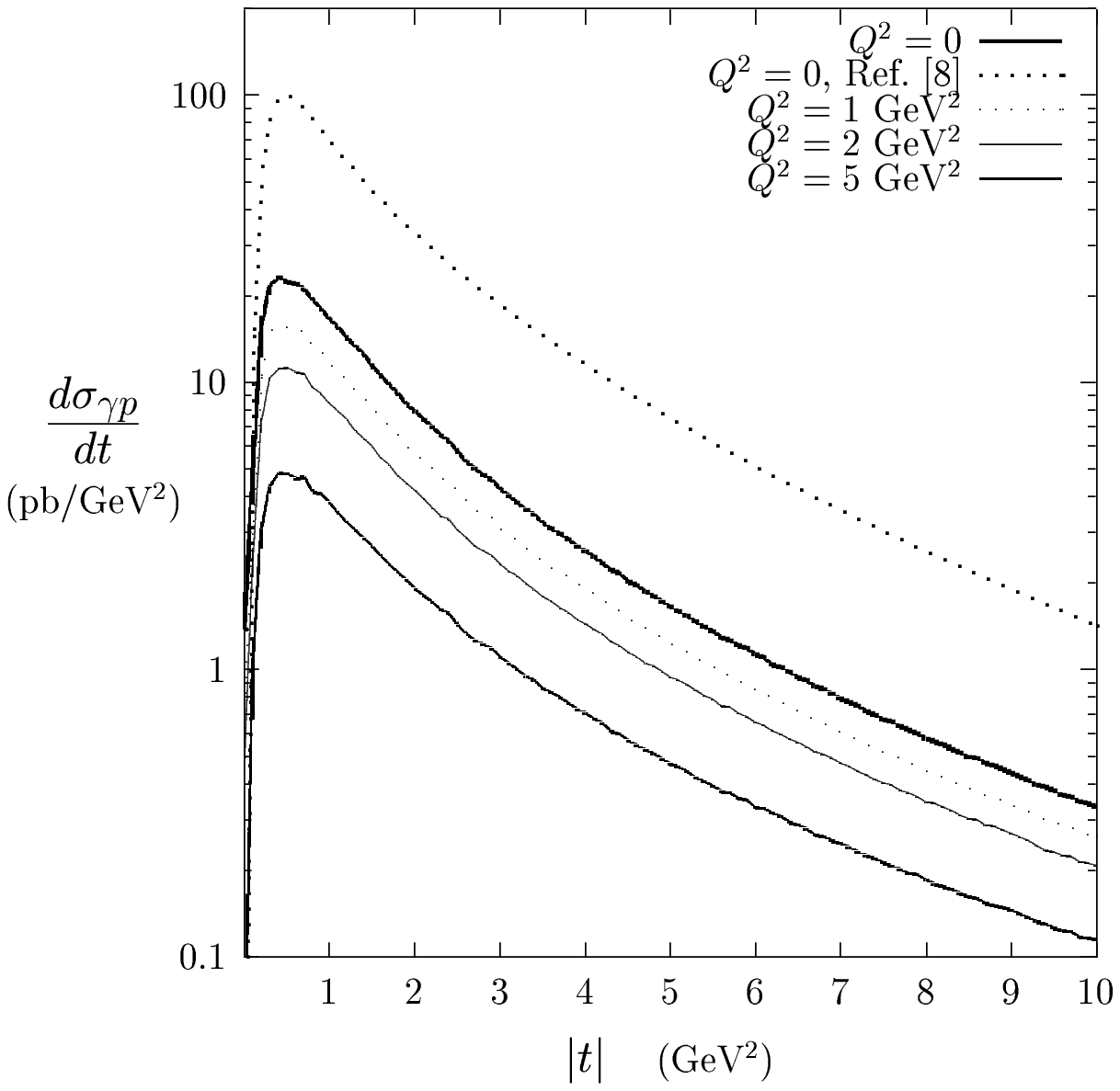}
\vskip.1in
\hspace{2cm}
(a)$\hspace{8cm}$(b)
\end{center}
\caption{\it (a) Elastic diffractive process $\gamma^*p\to M\,p$. (b) Differential cross section for  \hspace*{2cm}$\gamma^* p \to \eta_c\,p$ predicted in Ref. \cite{etac}.}
\end{figure}

One could expect that the largest cross section should be obtained in the photoproduction of 
light pseudoscalar meson M, i.e. for $M=\pi^0$. Due to  the absence in this case of a hard scale a fully non-perturbative description is necessary, such as  the stochastic vacuum model of the Heidelberg group which was quite succesfull in the description of processes involving pomeron exchanges. It turned out, however, that the predictions of this model for the diffractive photoproduction of $\pi^0$ \cite{pi1} has been disproved by  HERA measurements showing 
 much smaller values of the cross sections. A possible excuse for this disagreement is related to  peculiar properties of a pion as a Goldstone boson of the chiral symmetry of QCD \cite{pi2}.

\subsection{Pomeron-Odderon charge asymmetry}

The cross section for the diffractive production of mesons involves the square of the scattering amplitude with exchange of the odderon. In view of the presumed smallness  of odderon contribution it is desirable to have an observable which is linear in the amplitude which involves the odderon. Such a possibility is given by  the charge asymmetry measured in the diffractive electroproduction of $\pi^+ \pi^-$ meson pair as the one shown in Fig.~3a, 
$e(p_i, \lambda)\;\; N (p_N) \to e(p_f)\;\;\pi^+(p_+)\;\; \pi^-(p_-)\;\;
N^{\prime}(p_{N^{\prime}})$ \cite{POint1,POint2,POint3}. The state of $\pi^+ \pi^-$ meson pair is not an eigenstate of the C parity operator so the production mechanisms involves contributions from both, the pomeron and the odderon exchanges. 
The meson pair dynamics is governed by its generalized distribution amplitudes \cite{GDA} which take into account the final state interactions between mesons.
The charge asymmetry is defined as a ratio of the difference between integrated number of events with $\pi^+$ emitted  in direction of the polar angle $\theta$ determined in the cms of the pion pair and those emitted in direction of $\pi - \theta$ to the total number of events
\be
A(Q^{2},t,m_{2\pi }^{2},y,\alpha )=\frac{\sum\limits_{\lambda =+,-}\int \cos
\theta \,d\sigma (s,Q^{2},t,m_{2\pi }^{2},y,\alpha ,\theta ,\lambda )}{%
\sum\limits_{\lambda =+,-}\int d\sigma (s,Q^{2},t,m_{2\pi }^{2},y,\alpha
,\theta ,\lambda )}\,.
\ee
Here $d\sigma (s,Q^{2},t,m_{2\pi }^{2},y,\alpha ,\theta ,\lambda )$ is the cross section for production of meson pair with the invariant mass $m_{2\pi}$ from a
 photon with virtuality $Q^2$ carrying fraction $y$ of energy of the initial electron with helicity $\lambda$, see Fig.~3a. The charge asymmetry  $A(Q^{2},t,m_{2\pi }^{2},y,\alpha )$ defined above can be roughly treated as a "ratio"
 of the scattering amplitude with odderon exchange and the one with the pomeron exchange.
\begin{figure}[h]
\center{
\includegraphics[width=3.5in]{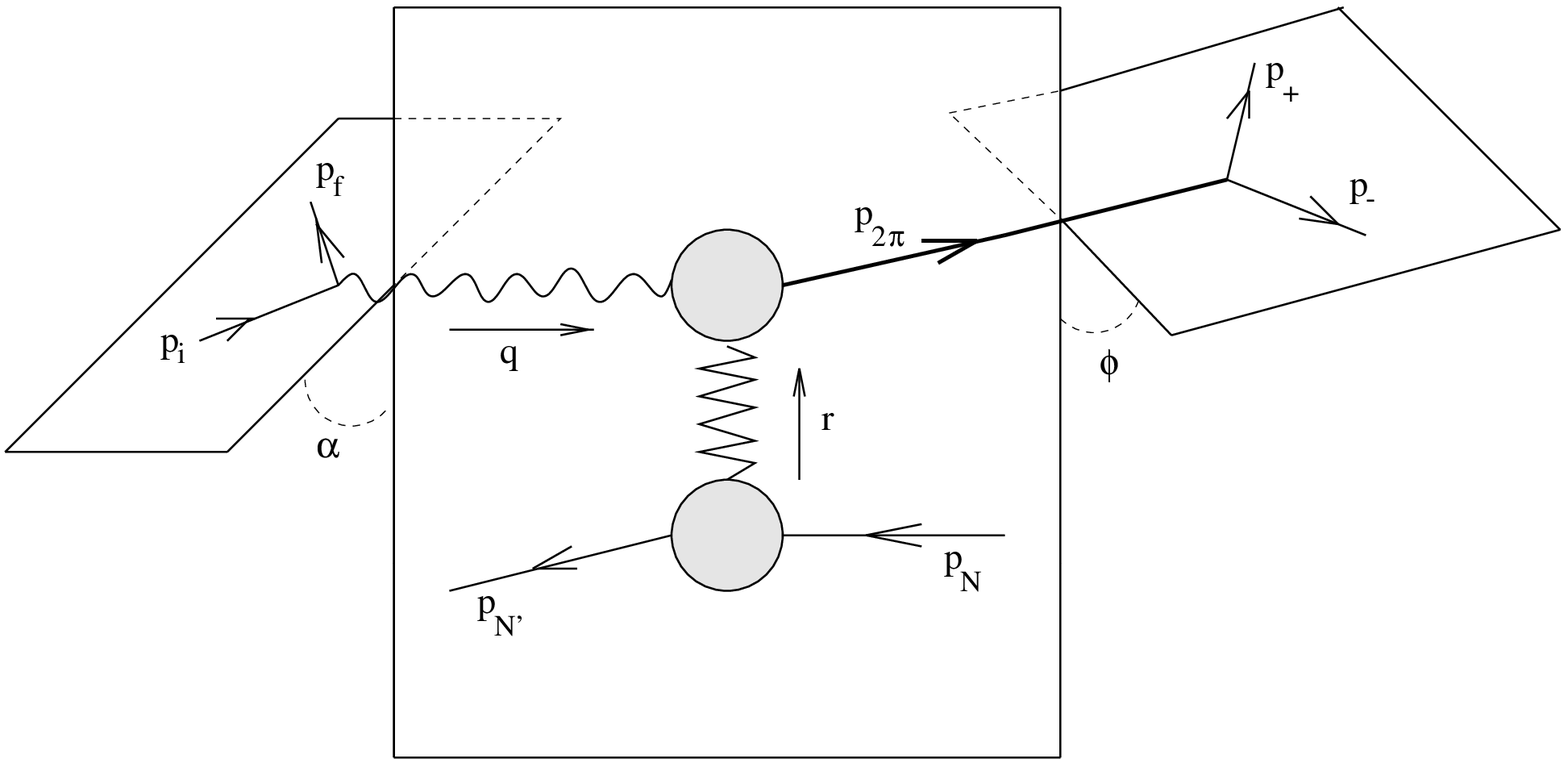}
\;\;\;\;\;\;
\includegraphics[width=2.4in]{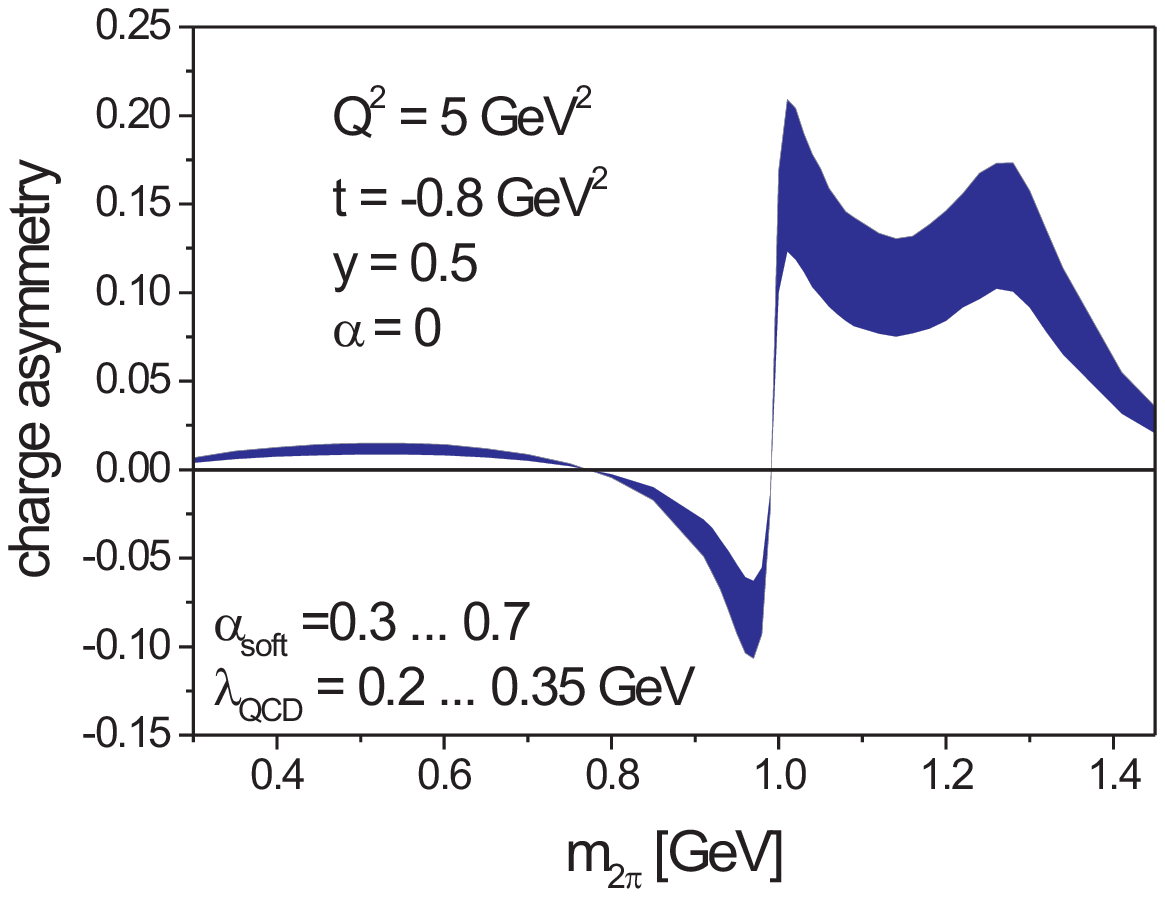}
\hspace{1.5cm}
(a)$\hspace{8cm}$(b)
\caption{(a) Kinematics of the electroproduction of two pions. (b) The charge asymmetry 
\hspace*{2cm} predicted in Ref.~\cite{POint2}  }
}
\end{figure}
The predicted charge asymmetry in Ref.~\cite{POint2} is shown in Fig.~3b. It is quite large in the region of $m_{2\pi }\approx 1.2\,$GeV corresponding to the $f_2$ meson mass. It would be highly desirable to perform study of the charge asymmetry based on  the H1 and ZEUS data, similarly as the one done by HERMES at much lower energies.

\section{Exclusive $J/\psi$ and $\Upsilon$ hadroproduction}

The new analysis of exprimental data on the 
 exclusive hadroproduction processes by the CDF collaboration \cite{tevatron} shows that
 these types of processes can be  objects of studies at the Tevatron and in the near future at the LHC.
 This opens a possibility to study the odderon effects in the exclusive hadroproduction of  
$J/\Psi$  and  $\Upsilon$ mesons (Fig.~4):\\
$pp\,(\bar{p})\,\rightarrow \, p' \, V \,p''\,(\bar{p}''\ )\;,\;\;\;\;\ \mbox{where}\;\;\;\;\; V=J/\psi,\, \Upsilon\;.
$
The
 production of a charmonium $V$, with
 the quantum numbers $J^{PC}=1^{-\ -}$, occurs in the above process as the result of a pomeron-odderon or pomeron-photon fusion \cite{ourwork}. 
 \begin{figure}[h]
\epsfxsize=3.5cm
\begin{center}
\epsffile{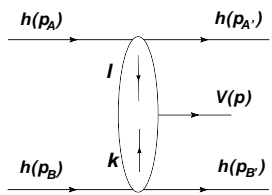}
\end{center}
\caption{\it Kinematics of the exclusive meson production in $pp$ ($p\bar p$) scattering.}
\end{figure} 
The lowest order contribution to the amplitude  
is illustrated by  Fig. 5, from which the diagrams (a,b) describe the pomeron-odderon fusion and (c,d)  
the photon-pomeron fusion.
\begin{figure}[h]
\begin{center}
{\large\bf a)}
\epsfxsize=2.8cm \epsffile{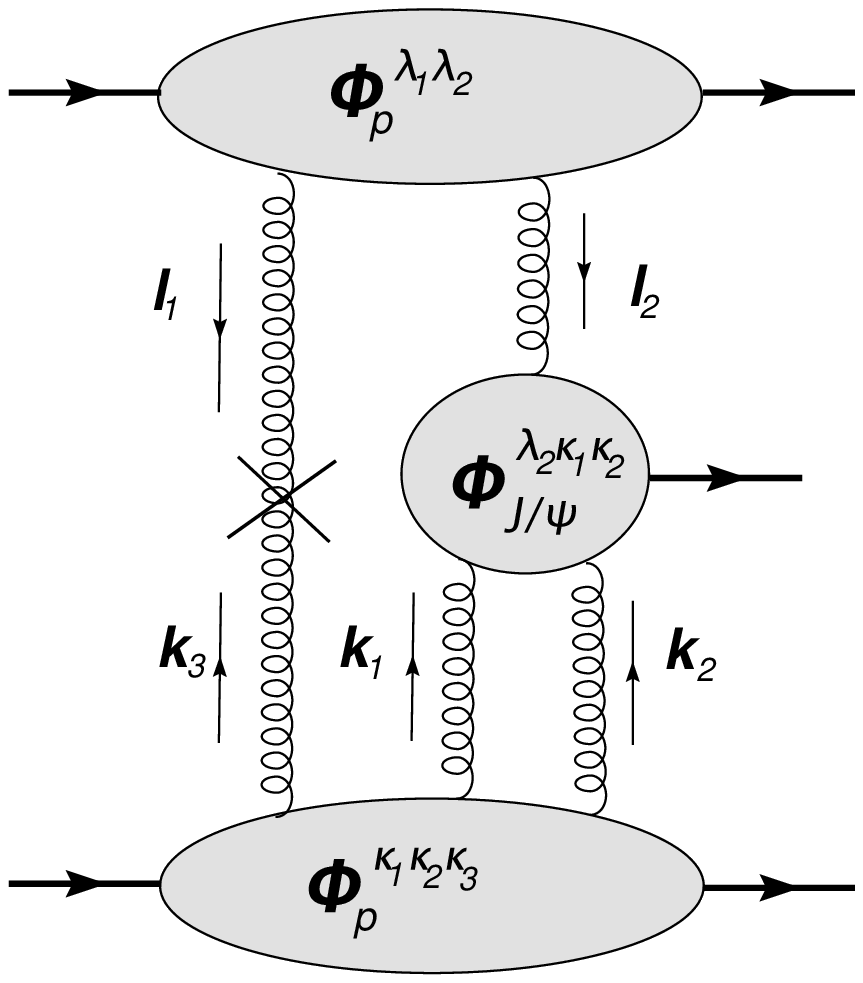} \hspace{0.3cm}
{\large\bf b)}
\epsfxsize=2.8cm \epsffile{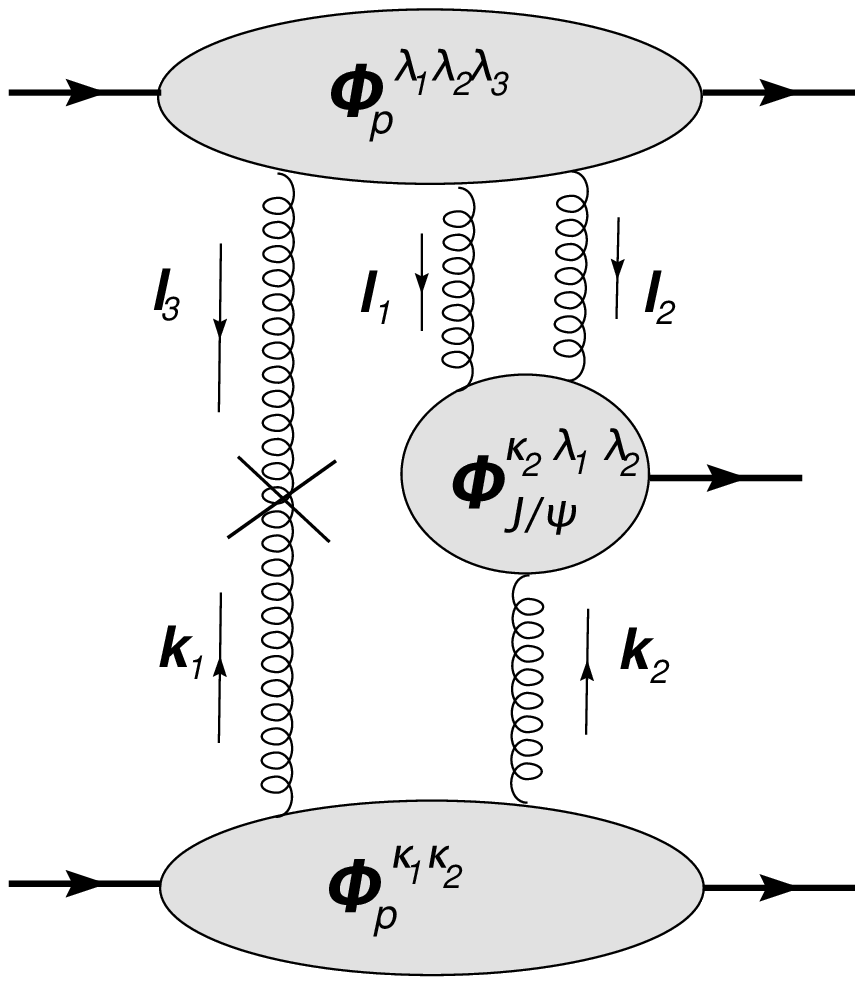} \hspace{0.3cm}
{\large\bf c)}
\epsfxsize=2.8cm \epsffile{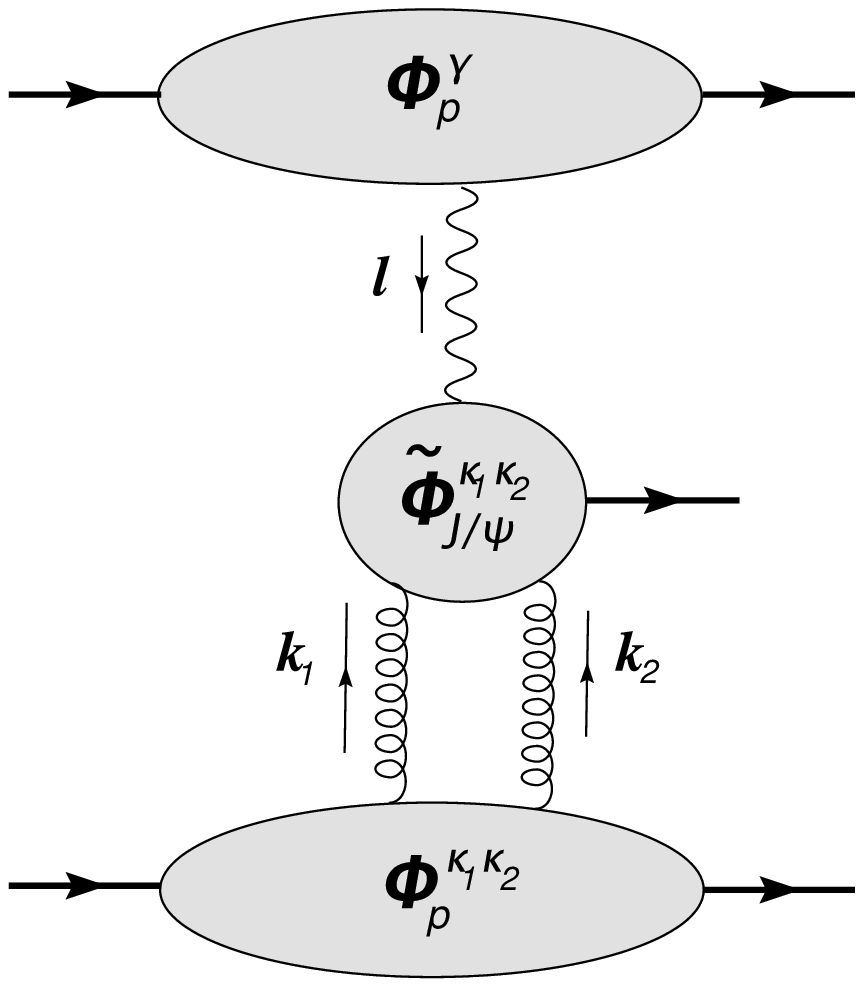} \hspace{0.3cm}
{\large\bf d)}
\epsfxsize=2.8cm \epsffile{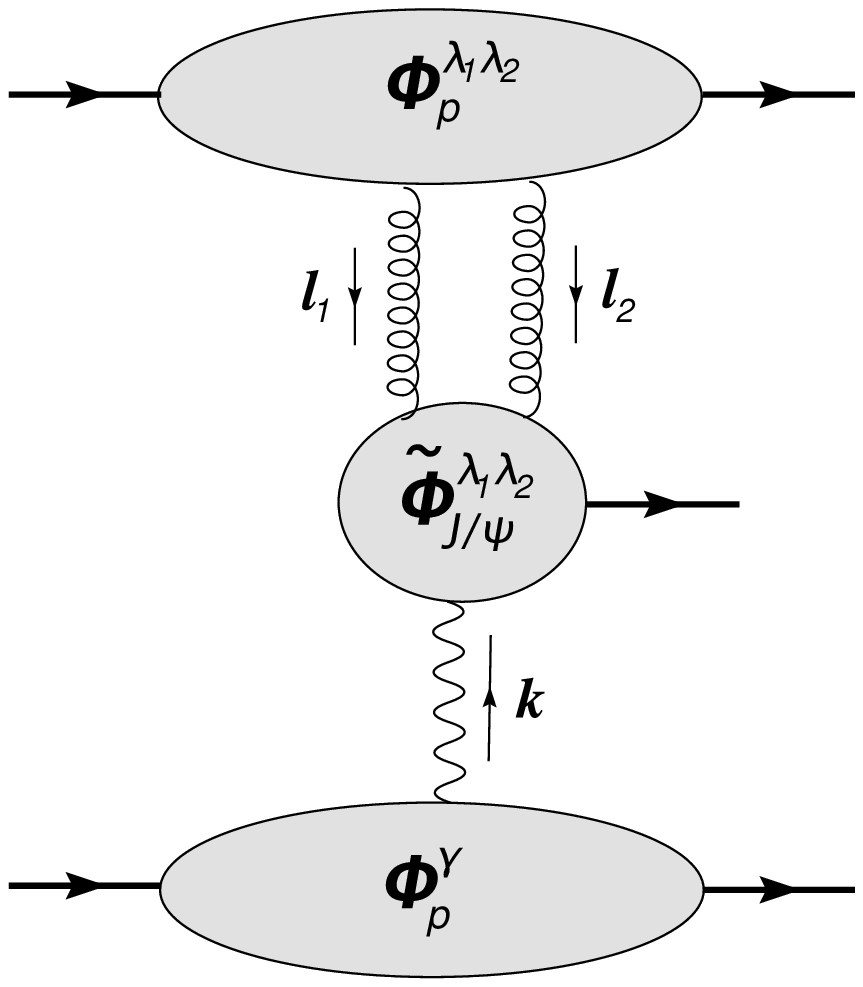} 
\end{center}
\caption{\it The lowest order diagrams contributiong to  the pomeron-odderon fusion (a,b) and  the pomeron--photon fusion (c,d)  for
 vector meson production.}
\end{figure}

The scattering amplitude written within the $k_\perp$-factorization approach is a convolution in transverse momenta of
$t-$channel fields. For instance,  the contribution of Fig. 5a  reads:
\bea
\label{impactA}
&& \hspace{-0.3cm}\cM_{P\;O}=
-is\;\frac{2\cdot3}{2!\,3!}\,\frac{4}{(2\pi)^8}\,\int \frac{d^2 \bl_1}{\bl_1^2}\,\frac{d^2 \bl_2}{\bl_2^2}\ \delta^2(\bl_1 + \bl_2 - \bl)\,\frac{d^2 \bk_1}{\bk_1^2}\,\frac{d^2 \bk_2}{\bk_2^2}\,\frac{d^2 {\bk}_3}{\bk_3^2}\ \delta^2(\bk_1+\bk_2+\bk_3-\bk)
\nonumber \\
&&
\times \delta^2(\bk_3+\bl_1)\,\bk_3^2\;\delta^{\lambda_1 \kappa_3}
\cdot\Phi^{\lambda_1 \lambda_2}_P(\bl_1,\bl_2)\cdot\Phi^{\kappa_1 \kappa_2 \kappa_3}_P(\bk_1,\bk_2,\bk_3)\cdot
\Phi_{J/\psi}^{\lambda_2 \kappa_1 \kappa_2}(\bl_2, \bk_1,  \bk_2)\;,
\eea
where $\Phi^{\lambda_1 \lambda_2}_P(\bl_1,\bl_2)$ and $\Phi^{\kappa_1 \kappa_2 \kappa_3}_P(\bk_1,\bk_2,\bk_3)$ 
are the impact factors describing the coupling of the pomeron and the odderon to  scattered hadrons, respectively,
whereas
$\Phi_{J/\psi}^{\lambda_2 \kappa_1 \kappa_2}(\bl_2, \bk_1,  \bk_2)$ is the effective $J/\psi$-meson production vertex. 
In writing this formula we used the Sudakov decomposition of momenta with respect to the incoming hadron momenta treated as two light-cone vectors and we used the euclidean notation for transverse vectors $l_\perp^2=- \,\bf l^2$.

\noindent
The proton impact factors are non-perturbative objects and we describe them within the Fukugita-Kwieci\'nski eikonal model \cite{Fukugita}. Their calculation is standard.
The derivation of the effective production vertex of $J/\psi$, $\Phi_{J/\psi}^{\lambda_2 \kappa_1 \kappa_2}(\bl_2, \bk_1,  \bk_2)$, in Eq.~(\ref{impactA}) is one of the main results of our study \cite{ourwork}. The charmonium is treated in the non-relativistic approximation and  the $\bar c c \to J/\psi$ vertex is described by
\be
\label{JPsivertex} 
\hspace*{-0.4cm} 
\langle \bar c \ c |J/\psi \rangle = \frac{g_{J/\psi}}{2}\ \hat \varepsilon^{\ *}(p)\left( p \cdot \gamma  + m_{J/\psi}  \right)\;,\;
m_{J/\psi}=2m_c\;,\;g_{J/\psi}=\sqrt{ \frac{3m_{J/\psi} \Gamma^{J/\psi}_{e^+e^-}}{16\pi \alpha_{em}^2 Q_c^2} }\;,\;Q_c=\frac{2}{3}\;.
\ee
with the coupling constant $g_{J/\psi}$ related to the electronic width $\Gamma^{J/\psi}_{e^+e^-}$ of the $J/\psi$.
The effective vertex $g+2g \to J/\psi$ is described by the sum of the contributions of the diagrams
in Fig.~6 which has the form 
\begin{figure}[t]
\centerline{
\epsfysize=4.0cm
\epsfxsize=9.0cm
\epsffile{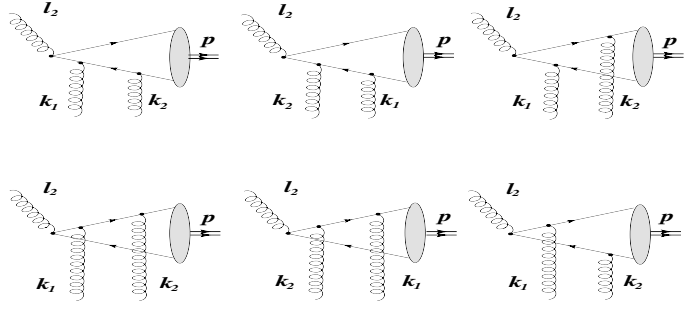} }
\caption{\it The six diagrams defining the effective vertex $g+2g \to J/\psi$.}
\end{figure}
\bea
\label{vertexPO}
&&\Phi_{J/\psi}^{\lambda_2 \kappa_1 \kappa_2}(\bl_2, \bk_1,  \bk_2) =
\alpha_s ^{3\over 2} \, 8\pi^{3\over 2}\; \frac{d^{\kappa_1\kappa_2\lambda_2}}{N_c}\;V_{J/\psi}(\bl_2, \bk_1,  \bk_2), 
\\
&& \hspace{-1.3cm}V_{J/\psi}(\bl_2, \bk_1,  \bk_2)=
4\pi m_c g_{J/\psi}\left[ - \frac{x_B \varepsilon^*\cdot p_B + \varepsilon^*\cdot l_{2\perp}}{\bl_2^2+(\bk_1+\bk_2)^2+4m_c^2} + \frac{\varepsilon^*\cdot l_{2\perp} +\varepsilon^*\cdot p_B\left( x_B - \frac{4\bk_1\cdot \bk_2}{sx_A} \right)}{\bl_2^2+(\bk_1-\bk_2)^2 +4m_c^2}  \right]\;,
\nonumber
\eea
where $x_{A/B}$ are the fractions of light-cone momenta $p_{A/B}$ carried by gluons in two $t-$channels.
In the numerical analysis we set $\alpha_s(m_c)=0.38$ and $\alpha_s(m_b)=0.21$.

\noindent
We describe the photon-pomeron fusion in Fig.~5c in a similar framework and obtain the expression 
\be
\label{impactC}
{\cal M}_{\gamma\, P}=
-\frac{1}{2!} \cdot s \cdot \frac{4}{(2\pi)^4\ \bl^2} \ \Phi^\gamma_P(\bl)\ \int \frac{d^2\bk_1}{\bk_1^2}
\frac{d^2 \bk_2}{\bk_2^2}\ \delta^2(\bk_1+\bk_2-\bk)\ \Phi^{\kappa_1\kappa_2}_P(\bk_1,\bk_2)\ \tilde \Phi^{\kappa_1\kappa_2}_{J/\psi}(\bl,\bk_1,\bk_2)\;,
\ee
where $\Phi^\gamma_P(\bl)$ is the phenomenological form-factor of the photon coupling to the proton chosen 
in consistent way with previously introduced impact factors within the Fukugita-Kwiecinski model.
The corresponding effective vertex
$\tilde \Phi^{\kappa_1\kappa_2}_{J/\psi}(\bl,\bk_1,\bk_2)$ is 
expressed through $V_{J/\psi}(\bl, \bk_1,  \bk_2)$ in Eq.~(\ref{vertexPO}) as
\be
\label{vertexGP}
\tilde \Phi^{\kappa_1\kappa_2}_{J/\psi}(\bl,\bk_1,\bk_2) = \, \alpha_s \,  eQ_c\, 8\pi\,
\frac{\delta^{\kappa_1\kappa_2}}{N_c}\,V_{J/\psi}(\bl, \bk_1,  \bk_2)\, .
\ee
The phases of the scattering amplitudes describing the two mechanisms of $J/\psi$-meson production differ by the factor $i=e^{i\pi/2}$. Consequently, they do not interfere and the cross section as a sum of two independent contributions: $|{\cal M}_{PO} ^{\mathrm{tot}}|^2 = |{\cal M}_{PO} +  {\cal M}_{OP}|^2$
and $|{\cal M}_{\gamma P} ^{\mathrm{tot}}|^2 = |{\cal M}_{\gamma P} +  {\cal M}_{P \gamma}|^2$.

\noindent
The realistic cross-sections are obtained from  the above expressions by taking into account phenomenological improvements, such as those related to the
BFKL evolution (which is very important for the pomeron exchange and which may be  omitted for the odderon exchange), the effects of soft rescatterings of hadrons, and the precise determination of the value of the model parameter $\bar \alpha_s$ in the impact factors.  We write the corrected cross-sections in the form
\be
\hspace*{-1cm}
\left.
{d \sigma^{\mathrm{corr}}_{PO} \over dy} 
\right|_{y=0}
\, = \, 
\bar\alpha_s^5\, S_{\mathrm{gap}}^2\, E(s,m_V) \,{d \sigma_{PO} \over dy},\;\;\;\;\;\;\;\;\;\left. {d \sigma^{\mathrm{corr}}_{\gamma P} \over dy}\right|_{y=0} \, = \, 
\bar\alpha_s^2\, E(s,m_V) \,{d \sigma_{\gamma P} \over dy},
\label{master}
\ee
where ${d \sigma_{PO/\gamma P}/ dy}$ are the cross sections determined  at
$\bar\alpha_s=1$.
The BFKL evolution for pomeron exchange  is taken into account by inclusion of the enhancement factor, which for the central production (i.e. for the rapidity $y = 0$) has the form 
$
E(s,m_V) = (x_0 \sqrt{s}/m_V)^{2\lambda}.
$
Here, $x_0$ is
the maximal fraction of incoming hadron momenta exchanged in the $t-$channels (or the initial condition for the BFKL evolution) and it is set to $x_0=0.1$.
The effective pomeron intercept $\lambda$ is determined by HERA data, see \cite{ourwork}. 
The gap surviving factor $S_{\mathrm{gap}}^2$ for the exclusive production via the pomeron-odderon fusion is fixed by the results of the Durham two channel eikonal model 
\cite{two-channel}: $S_{\mathrm{gap}}^2=0.05$ for the exclusive production at the Tevatron and $S_{\mathrm{gap}}^2=0.03$ for LHC. In the case of production from the photon-pomeron fusion, 
$S_{\mathrm{gap}}^2=1$ \cite{KMR-phot}.

\noindent
The available estimates of the effective strong coupling constant 
$\bar\alpha_s$ in the Fukugita--Kwieci\'{n}ski model 
have a rather large spread: from $\bar\alpha_s\approx 1$  \cite{Fukugita}, through $\bar\alpha_s\approx 0.6-0.7$ determined from the HERA data 
 to $\bar\alpha_s\approx 0.3$ determined from data on  elastic
$pp$ and $p\bar p$ scattering \cite{Ewerz-coupling}. 
This led us to introduce  three scenarios which differ by the values of $\bar\alpha_s$ and of $S_{\mathrm{gap}}^2$.
In the {\em optimistic scenario} we  use a large value of the
coupling, $\bar\alpha_s = 1$, combined with the gap survival factors 
obtained in the Durham two-channel eikonal model. We believe that the best estimates should follow 
from the {\em central scenario} defined by 
$\bar\alpha_s=0.75$, and Durham group estimates $S_{\mathrm{gap}}^2 =0.05$ ($S_{\mathrm{gap}}^2=0.03$) 
at the Tevatron (LHC). The {\em pessimistic scenario} is defined by $\bar\alpha_s=0.3$ and $S_{\mathrm{gap}}^2=1$.

\begin{table}[t]
\begin{center}%
\label{tab2}
\begin{tabular}[c]{|c|c|c|c|c|}\hline\hline
$d\sigma^{\mathrm{corr}}/dy$ & 
\multicolumn{2}{c|}{$ J/\psi$} &
\multicolumn{2}{c|}{$\Upsilon$} \\ \cline{2-5}
 &  odderon & photon & odderon & photon \\ \hline
Tevatron &
0.3--1.3--5~nb & 
0.8--5--9~nb &  
0.7--4--15~pb & 
0.8--5--9~pb  \\
LHC      & 
0.3--0.9--4~nb & 
2.4--15--27~nb  & 
1.7--5--21~pb & 
5--31--55~pb\\\hline \hline
\end{tabular}
\end{center}
\caption{
 The phenomenologically corrected cross sections $d\sigma^{\mathrm{corr}} /dy|_{y=0}$ 
 for the exclusive $J/\psi$ and $\Upsilon$ production 
in $pp$ and $p\bar p$ collisions by the pomeron--odderon  
fusion, and analogous cross sections  
$d\sigma^{\mathrm{corr}} _{\gamma} /dy|_{y=0}$ for the photon contribution
 for the 
pessimistic--central--optimistic scenarios.}
\end{table}

\noindent
Table 1 shows our predictions for the phenomenologically improved cross sections in all three scenarios. 
Their magnitudes justify our hope that the hadroproduction process (\ref{genprocess}) is a promising  subject of experimental 
study in the near future at the Tevatron and at the LHC \cite{alice}. The encouraging feature of our results is due to the fact, that
the measurement of the $t_i$ dependence of the cross section partially permits filtering out the $\gamma\ P$ contributions and to uncover  the $PO$ ones.

\section*{Acknowledgments}

I acknowledge the common research and discussions with A. Bzdak, J-R. Cudell and L. Motyka. This work is supported by the Polish Grant 1 P03B 028 28.


\end{document}